\documentclass[a4paper,10pt,eqno]{article}
\usepackage{latexsym,amssymb,amsfonts,amsmath}
\usepackage{amsthm}
\usepackage[dvips]{graphicx}
\usepackage{color}
\newtheorem{thm}{Theorem}[section]
\newtheorem*{thm*}{Theorem}
\newtheorem{lem}[thm]{Lemma}
\newtheorem*{lem*}{Lemma}

\newtheorem*{cor*}{Corollary}

\newtheorem{pro}[thm]{Proposition}

\newcommand{\ZM}{\mathbb{Z}}
\newcommand{\ZMP}{\mathbb{Z}_{+}}
\newcommand{\NM}{\mathbb{N}}
\newcommand{\CM}{\mathbb{C}}

\newcommand{\PM}{\mathbb{P}}

\newcommand{\QC}{\mathcal{Q}}
\newcommand{\ket}[1]{|#1\rangle}
\newcommand{\bra}[1]{\langle #1|}
\newcommand{\braket}[2]{\langle #1 | #2 \rangle}

\newcommand{\lr}[1]{\left( #1\ \right)}

\title{{\Large {\bf Periodicity for the Fourier quantum walk on regular graphs}}}
\author{
{\small Kei Saito\footnote{saito-kei-nb@ynu.jp},}\\
{\scriptsize  Department of Applied Mathematics, Faculty of Engineering Science, Yokohama National University}\\
{\scriptsize \footnotesize\it 79-5 Tokiwadai, Hodogaya, Yokohama, 240-8501, Japan}\\
}

\date{\empty}
\begin{document}
\maketitle

\par\noindent
\begin{small}
\par\noindent
{\bf Abstract}.
 Quantum walks determined by the coin operator on graphs have been intensively studied.
The typical examples of coin operator are the Grover and Fourier matrices.
The periodicity of the Grover walk is well investigated.
However, the corresponding result on the Fourier walk is not known.
In this paper, we present a necessary condition for the Fourier walk on regular graphs to have the finite period.
As an application of our result, we show that the Fourier walks do not have any finite period for some classes of regular graphs such as complete graphs, cycle graphs with selfloops, and hypercubes.

\footnote[0]{
{\it Abbr. title:} Periodicity for the Fourier quantum walk on regular graphs
}
\end{small}

\setcounter{equation}{0}

\section{Introduction}
The discrete-time quantum walk (QW) is a quantum dynamics defined as a quantization of the classical random walk \cite{QW1,QW2}.
QWs have been studied from various research fields, e.g., quantum physics, information sciences.
Some recent reviews and books are \cite{review1,review2,review3,review4}.
Especially, QWs on graphs are widely investigated and expected to be an application of quantum searching algorithm.
It is known that some searching algorithm using QWs achieved quadratic speed up compared with classical algorithm. \cite{alg1,alg2,alg3,review4} 

In this paper, we focus on the {\it periodicity} of the QW on regular graphs as a main topic.
The dynamics of the QW is determined by the time evolution operator $U$.
This operator is given by the product of two unitary matrices, shift operator and coin operator.
Here, if there exists a positive integer $T(<\infty)$ such that $U^T$ is identity, then we say that this QW is periodic with the period $T$.
That is, the quantum state recurs periodically when the QW has the finite period.
As typical models of QWs, there are the Grover QW (GQW) and the Fourier QW (FQW).
The GQW is a special case of the Szegedy's QW which defined as a quantum extension of the Markov chain on graphs \cite{Sze}.
The coin operator of the GQW is given by the Grover matrix and the spectrum of the time evolution operator is derived from that of the corresponding random walk.
Hence, the GQW is well studied and its periodicity is clarified for some finite graphs, e.g., complete graphs, cycle graphs, the generalized Bethe trees \cite{period1, period6, period7, period8}.
Here, the FQW is defined by taking its coin operator as the Fourier matrix.
In contrast to the GQW, any rigorous results of the FQW except the Hadamard walk, which is a special model of the FQW, is not known.
In fact, the period $T$ of the Hadamard walk on the cycle graph with number of vertex $N(\geq 2)$ is completely obtained as follows \cite{period4,period5}.
\begin{align*}
T=
\begin{cases}
2 \quad &(N=2)\\
8 \quad &(N=4)\\
24 \quad &(N=8)\\
\infty \quad &(N\neq 2,4,8)\\
\end{cases}
.
\end{align*}
Here, $T=\infty$ means the QW is not periodic.
Our purpose is to specify the structure of graph on which the FQW is periodic.
In the present paper, we give a necessary condition for the FQW on the regular graph to have the finite period.
This result shows that if the FQW has a finite period, then a strong restriction is imposed on the construction of the graph (see Theorem \ref{mainthm}).
Let $\PM$ and $\NM$ denote the sets of all prime numbers and natural numbers, respectively.
As an application of our result, we prove that the period is infinite for the FQWs on three typical regular graphs: complete graph $K_N$ with $N=p^n+1$, cycle graph with selfloops $C_N^{(l)}$ with $N\neq 3^n\ (N>2)$, and hypercube $\QC_d$ with dimension $d=p^n\ (p\neq 2)$, where $p\in\PM$ and $n\in\NM$.

The rest of this paper is organized as follows.
In Sect. 2, we introduce the definition of the QW on graphs.
Section 3 is devoted to our main result and its proof. 
In Sect. 4, we apply our result to some typical regular graphs. 
Section 5 deals with summary. 

\section{Definition of FQW on regular graphs}
First of all, we will introduce the definition of the QW on the finite $k$-regular multigraph $G=(V(G),E(G))$ in this section. 
Here, $V(G)$ and $E(G)$ denote the sets of vertices and edges of $G$, respectively.
Throughout this paper, we write $V(G)=\{0,1,\ldots,N-1\}$, where $N$ is the number of vertices of $G$.
Furthermore, each element of $E(G)$ is expressed as $(xy)_{m(xy)}\ (x,y\in V(G))$, where $m(xy)$ denotes the multiplicity of the edge.
In this paper, we use $m$ as an abbreviation of $m(xy)$.
In order to define the dynamics of the QW, we give the set of symmetric arcs $D(G)=\{(x,y)_{m}, (y,x)_{m}\,|\,(xy)_{m}\in E(G)\}$.
We let the initial and terminal vertices of $e=(x,y)_{m}\in D(G)$ as $o(e)=x$ and $t(e)=y$, respectively.
\\
Let
\begin{align*}
\Psi_t=\sum_{s=0}^{k-1}\sum_{x=0}^{N-1}\psi_{t,s}(x)\ket{x,s}\in\left(\CM^{k}\right)^{N}
\end{align*}
be the quantum state at time $t$, where each $\psi_{t,s}(x)$ is a complex component of $\Psi_t$.
More precisely, we put
\begin{align*}
\Psi_t=
{}^T
\begin{bmatrix}
\begin{bmatrix} \psi_{t,0}(0) \\ \psi_{t,1}(0) \\ \vdots \\ \psi_{t,k-1}(0)\end{bmatrix}
&
\begin{bmatrix} \psi_{t,0}(1) \\ \psi_{t,1}(1) \\ \vdots \\ \psi_{t,k-1}(1)\end{bmatrix}
&
\cdots
&
\begin{bmatrix} \psi_{t,0}(N-1) \\ \psi_{t,1}(N-1) \\ \vdots \\ \psi_{t,k-1}(N-1)\end{bmatrix}
\end{bmatrix},
\end{align*}
where $T$ is the transpose operator.
The time evolution of the QW is defined as $U=S(I_{N}\otimes C)$ by using two matrices $S$ and $C$ (defined later),  where $I_n$ means the $n\times n$ identity matrix.
That is, the time evolution of the QW is given as follows.
\begin{align*}
\Psi_t=U^t\Psi_0.
\end{align*}
From now on, we will explain the definitions of the matrices $S$ and $C$.
Firstly, $S$ is given by a labelling of arcs $\sigma : D(G)\to [k]$ satisfying $\sigma(e)\neq\sigma(e')$ when $o(e)=o(e')$ or $t(e)=t(e')$, and a bijection $\pi : [k]\to[k]$, where $[k]=\{0,1,\ldots ,k-1\}$.
$\sigma_x$ is a restriction of $\sigma$ to the arcs whose initial vertex is $x$.
Remark that $\sigma_x$ is a bijection. 
By using $\sigma$ and $\pi$, $S$ is defined as
\begin{align*}
S\ket{x,s}=\ket{t(\sigma_x^{-1}(s)),\pi(s)}.
\end{align*}
Then, $S$ is given as a permutation and called {\it shift operator}.
Moreover, $\pi$ governs the shift of quantum states.
In other words, $\pi$ flips the direction of the quantum walker.
For example, if $\pi$ is the identity map, then it is called ``moving shift": this shift operator does not change the direction of the quantum walker after every move.
When $\pi(\sigma_x(x,y)_m)=\sigma_y(y,x)_m$ for each $x,y\in V(G)$, it is called ``flip-flop shift": this shift operator reverses the direction of the quantum walker after every move.

Secondly, $C$ is defined by a $k\times k$ unitary matrix and called {\it coin operator} .
Here, we will rewrite the time evolution of the QW focusing on each vertex.
Then, as in the following argument, the weight of each path on which the walker moves can be written by the division of the coin operator $P_{a:b}=\ket{b}\bra{a}C$.
Let
\begin{align*}
\Psi_t(x)=\sum_{s=0}^{k-1}\psi_{t,s}(x)\ket{s}\in\CM^k
\end{align*}
be an amplitude at time $t$ on the vertex $x$.
The quantum state is expressed as follows.
\begin{align*}
\Psi_t=\sum_{x=0}^{N-1}\ket{x}\otimes\Psi_t(x).
\end{align*}
By using above notation, the time evolution is written as
\begin{align*}
S\lr{I_N\otimes C}\Psi_t&=S\sum_{x=0}^{N-1}\ket{x}\otimes C\Psi_t(x)
\\
&=S\sum_{x=0}^{N-1}\sum_{s=0}^{k-1}\psi_{t,s}(x)\ket{x}\otimes C\ket{s}
\\
&=S\sum_{x=0}^{N-1}\sum_{s=0}^{k-1}\sum_{j=0}^{k-1}\psi_{t,s}(x)c_{j,s}\ket{x,j}
\\
&=\sum_{x=0}^{N-1}\sum_{s=0}^{k-1}\sum_{j=0}^{k-1}\psi_{t,s}(x)c_{j,s}\ket{t(\sigma_x^{-1}(j)),\pi(j)},
\end{align*}
where $C=\lr{c_{u,v}}_{u,v=0,1,\ldots ,k-1}$.
Hence, the amplitude at time $t+1$ on the vertex $y$ is given as
\begin{align*}
\Psi_{t+1}(y)&=\sum_{x,j\,|\,t(\sigma_x^{-1}(j))=y}\sum_{s=0}^{k-1}\psi_{t,s}(x)c_{j,s}\ket{\pi(j)}
\\[+4pt]
&=\sum_{x,j\, | \, t(\sigma_x^{-1}(j))=y}\ket{\pi(j)}\bra{j}C\Psi_t(x)
\\[+4pt]
&=\sum_{x,j\, |\, t(\sigma_x^{-1}(j))=y}P_{\pi(j):j}\Psi_t(x).
\end{align*}
Here, the second equality is obtained by $\sum_{s=0}^{k-1}\psi_{t,s}(x)c_{j,s}=\bra{j}C\Psi_t(x)$.
When the graph has no multiedges, the vertex $x$ satisfying $t(\sigma_x^{-1}(j))=y$ is uniquely determined for fixed $j$.
Therefore, the time evolution is expressed as follows.
\begin{align}
\label{tev}
\Psi_{t+1}(y)=\sum_{x\,|\,x\sim y}P_{\pi(\sigma(x,y)):\sigma(x,y)}\Psi_t(x),
\end{align}
where $x\sim y$ means $xy\in E(G)$.
We should remark that the same argument holds for multigraphs.
In such a case, the weight of the walker hops to vertex $x$ to $y$ becomes sum of each $P_{\pi(\sigma(x,y)_m):\sigma(x,y)_m}$.
For any $t\in\ZM_+$, Eq. (\ref{tev}) gives a vertex-based expression of $U^t$ as
\begin{align*}
U^t=
\left(
\Xi_t(x,y)
\right)_{x,y=0,1,\ldots,N-1},
\end{align*}
where $\ZM_+=\{0,1,2,\ldots\}$ and $\Xi_t(x,y)$ is a $k\times k$ matrix defined by the sum of weight of possible paths from $y$ to $x$ at step $t$.
Furthermore, $\Xi_1(x,y)$ is composed by a linear combination of the division of coin operator $P_{\pi(\sigma(x,y)):\sigma(x,y)}$ whose coefficient is 0 or 1.
Especially, if $xy$ does not have multiplicity, then $\Xi_1(x,y)$ corresponds to $P_{\pi(\sigma(x,y)):\sigma(x,y)}$.

In summary, the dynamics of the QW on regular graphs is defined by the following four elements:
\begin{itemize}
\item $k$-regular multigraph $G=(V(G),E(G))$
\item initial state $\Psi_0$
\item shift operator $S$
\begin{itemize} 
\item labelling $\sigma$ : $D(G)\to [k]$ \quad ($\sigma(e)\neq\sigma(e')$ when $o(e)=o(e')$ or $t(e)=t(e')$)
\item shift of quantum states $\pi$ : $[k]\to[k]$ \quad (bijection)
\end{itemize}
\item coin operator $C$ : $k\times k$ unitary matrix
\end{itemize}
Here, we give an example.
\\[+5pt]
{\bf Example:}
\\
For the cycle graph $C_N\,(N>2)$ with $V(C_N)=[N]$ and $D(C_N)=\{(x,y)\ |\ y\equiv x\pm 1 \mod\ N\}$, the time evolution of the moving shift QW with $\sigma((x,y))= \begin{cases}0\ \  (y\equiv x-1 \mod N) \\ 1\ \ (y\equiv x+1 \mod N)\end{cases}$ is given by
\begin{align*}
U=
\begin{bmatrix}
O & P & O & \cdots & O & Q\\
Q & O & P & \ddots & O & O\\[-5pt]
O & Q & O & \ddots & \ddots & \vdots\\[-5pt]
\vdots &\ddots & \ddots &\ddots &\ddots & \vdots\\[-5pt]
O & O & \cdots &\ddots &O & P\\
P & O & \cdots & \cdots & Q & O
\end{bmatrix},
\end{align*}
where $P=P_{0:0},\ Q=P_{1:1},$ and $O$ is the $2\times 2$ zero matrix.\\[+5pt]
\indent
Here, we introduce the FQW whose coin operator is given by the following Fourier matrix:
\begin{align*}
F(k)=(\omega^{uv}/\sqrt{k})_{u,v=0,1,\ldots , k-1},
\end{align*}
where $\omega=e^{\frac{2\pi}{k}i}$.
The square of the Fourier matrix is calculated as $(F(k)^2)_{u,v=0,1,\ldots ,k-1}=\sum_{j=0}^{k-1}\omega^{j(u+v)}/k$.
Then, each component of $F(k)^2$ equals to $1\,(u+v\equiv 0 \ {\rm mod}\ k)$, or $0\, (u+v\not\equiv 0 \ {\rm mod}\ k)$.
Hence, $F(k)^4$ becomes identity matrix $I_k$.
Our purpose of this study is to analyse the {\it periodicity} of the FQW on regular graphs.
The QW has a period $T$ if and only if $T=\min\mathcal{N}$, where
\begin{align*}
\mathcal{N}=
\{
n\geq 1 :
U
^n
=
I_{k|V(G)|}
\}.
\end{align*}
We remark that if $\mathcal{N}=\emptyset$, then $T=\infty$ and we say that this QW is not periodic.
When the FQW has a finite period $T$, our main result on Sect.3 gives a strong restriction to the construction of graph.
Moreover, when all initial states of each vertex do not depend on the position, the time evolution is written as $\Psi_t=(I_{N}\otimes \left( F(k) \right)^t )\Psi_0$.
Then, the following proposition holds, since $(F(k))^4=I_k$, and $\Psi_T=\Psi_0$ if the FQW has a finite period $T$.
\vspace*{12pt}
\begin{pro}
\label{prop1}
For the FQW on regular graphs, if this QW has a finite period $T$, then we have
\begin{align*}
T\equiv 0\ \rm mod \ 4.
\end{align*}
\end{pro}

\section{Main result and its proof}
In this section, we will present the following main result.
\vspace*{12pt}
\begin{thm}
\label{mainthm}
For $k=p^n$ with $p\in\PM$ and an $n\in\NM$,
if the FQW on a $k$-regular graph has a period $T(<\infty)$, then the following relation holds.
\begin{align*}
W_T(x,y) \equiv 0 \ {\rm mod}\ p \quad (x,y\in V(G)),
\end{align*}
where $W_T(x,y)$ is the number of possible paths from $y$ to $x$ at $T$ step.  
\end{thm}
\vspace*{12pt}
This theorem shows that if the FQW has a finite period, then a strong restriction on $W_T(x,y)$ is imposed .
By using this result, we prove that the period is infinite for the FQWs on some class of three typical regular graphs, i.e., complete graph, cycle graph with selfloops, and hypercube (see Section 4).

\noindent
{\bf Proof:}\quad
First, we will present a key relation of this proof as follows.
\begin{align*}
P_{a:b}P_{a':b'}=\ket{b}\bra{a}F(k)\ket{b'}\bra{a'}F(k)=\frac{\omega^{ab'}}{\sqrt{k}}\ket{b}\bra{a'}F(k)=\frac{\omega^{ab'}}{\sqrt{k}}P_{a':b}.
\end{align*}
Here, the second equality is obtained by
\begin{align*}
\bra{a}F(k)\ket{b'}=\frac{\omega^{ab'}}{\sqrt{k}}.
\end{align*}
The third equality is given by definition of $P_{a:b}$.
Noting that $\Xi_t(x,y)$ can be written by a linear combination of $P_{a:b}$, we see that there exists $\tilde{p}_{a:b:c}(x,y,t)\in\ZM_+$ such that
\begin{align}
\label{til}
\Xi_t(x,y)=
\left(
\frac{1}{\sqrt{k}}
\right)^{t-1}
\sum_{a,b,c=0}^{k-1}
\omega^{c}
\tilde{p}_{a:b:c}(x,y,t)
P_{a:b}.
\end{align}
We should remark that 
\begin{align}
\label{sump}
\sum_{a,b,c=0}^{k-1}\tilde{p}_{a:b:c}(x,y,t)=W_t(x,y),
\end{align}
since $\tilde{p}_{a:b:c}(x,y,t)$ is given by the each path from $y$ to $x$ at $t$ step.
In order to explain $\tilde{p}_{a:b:c}(x,y,t)$ in the right hand side of Eq. (\ref{til}), from now on, we consider a moving shift FQW on $C_3$ with self loops, whose time evolution is determined by
\begin{align*}
U=
\begin{bmatrix}
R & P & Q\\
Q & R & P\\
P & Q & R
\end{bmatrix}
,
\end{align*}
where $P=P_{0:0},\ R=P_{1:1},\ Q=P_{2:2}$.
In particular, if we take $t=3$ and $x=y$, then we have
\begin{align*}
\Xi_{3}(x,x)=P^3+PRQ+PQR+R^3+RPQ+RQP+Q^3+QPR+QRP.
\end{align*}
Moreover, we compute
\begin{align*}
&P^3=
\left(
\frac{1}{\sqrt{3}}
\right)^2
\omega^{0}
P_{0:0},
\quad
PRQ=
\left(
\frac{1}{\sqrt{3}}
\right)^2
\omega^{2}
P_{2:0},
\quad
PQR=
\left(
\frac{1}{\sqrt{3}}
\right)^2
\omega^{2}
P_{1:0},
\quad
\\
&R^3=
\left(
\frac{1}{\sqrt{3}}
\right)^2
\omega^{2}
P_{1:1},
\quad
RPQ=
\left(
\frac{1}{\sqrt{3}}
\right)^2
\omega^{0}
P_{2:1},
\quad
RQP=
\left(
\frac{1}{\sqrt{3}}
\right)^2
\omega^{2}
P_{0:1},
\\
&Q^3=
\left(
\frac{1}{\sqrt{3}}
\right)^2
\omega^{2}
P_{2:2},
\quad
QPR=
\left(
\frac{1}{\sqrt{3}}
\right)^2
\omega^{0}
P_{1:2},
\quad
QRP=
\left(
\frac{1}{\sqrt{3}}
\right)^2
\omega^{2}
P_{0:2}.
\end{align*}
Therefore,
\begin{align*}
\tilde{p}_{a:b:c}(x,x,3)=
\begin{cases}
1\qquad \text{if }(a,b,c)\in \Omega
\\
0\qquad \text{otherwise}
\end{cases},
\end{align*}
where $\Omega=\{(0,0,0),\ (0,1,2),\ (0,2,2),\ (1,0,2),\ (1,1,2),\ (1,2,0),\ (2,0,2),\ (2,1,0)\}$.
In a similar way, we obtain Eq.~(\ref{til}) for general regular multigraphs.

Furthermore, Eq. (\ref{til}) implies that each component of $\Xi_t(x,y)$ is expressed as
\begin{align*}
\left(
\Xi_t(x,y)
\right)_{u,v}
=
\left(
\frac{1}{\sqrt{k}}
\right)
^{t}
\sum_{a=0}^{k-1}
\sum_{c=0}^{k-1}
\omega^{c+av}
\tilde{p}_{a:u:c}(x,y,t).
\end{align*}
Summing over all components of the first column of $\Xi_t(x,y)$, we get the following key equation of this proof.
\begin{align}
\label{key1}
\sum_{u=0}^{k-1}
\left(
\Xi_t(x,y)
\right)_{u,0}
=
\left(
\frac{1}{\sqrt{k}}
\right)
^{t}
\sum_{a,b=0}^{k-1}\sum_{c=0}^{k-1}\omega^{c}
\tilde{p}_{a:b:c}(x,y,t)
=
\left(
\frac{1}{\sqrt{k}}
\right)
^{t}
\sum_{c=0}^{k-1}\omega^{c}
\left(
\sum_{a,b=0}^{k-1}
\tilde{p}_{a:b:c}(x,y,t)
\right).
\end{align}
Here, if the QW has a period $T\, (<\infty)$, then $\Xi_T(x,y)=I_{k}\ (x=y)$ or $=O\ (x\neq y)$, since  $U^T$ becomes identity.
Hence, Eq. (\ref{key1}) gives
\begin{align}
\label{key2}
\sum_{c=0}^{k-1}\omega^{c}
\left(
\sum_{a,b=0}^{k-1}
\tilde{p}_{a:b:c}(x,y,T)
\right)
=
\begin{cases}
\left(
\sqrt{k}
\right)
^{T}
\quad
(x=y)
\\[+5pt]
\quad
0
\hspace{1.05cm}
(x\neq y)
\end{cases}
.
\end{align}
We should remark that Proposition \ref{prop1} guarantees $\left(
\sqrt{k}
\right)
^{T}\in\NM$.
Noting Eq.~(\ref{sump}), the desired conclusion is given by applying the following Proposition \ref{sumq} to Eq.~(\ref{key2}).
\vspace*{12pt}
\begin{pro}
\label{sumq}
Put $k=p^n$ with $p\in\PM$ and an $n\in\NM$.
For $q_j\in\ZM_+\ (j=0,1,\ldots ,k-1)$ satisfying $\sum_{j=0}^{k-1}q_j\omega^{j}=0$, the following relation holds.
\begin{align*}
\sum_{j=0}^{k-1}q_j\equiv 0\ {\rm mod\ }p. 
\end{align*}
\end{pro}

In order to prove this proposition, we will explain the cyclotomic polynomial $\Phi_m(x)$ given as
\begin{align*}
\Phi_m(x)=\prod_{\substack{1\leq j\leq m \\ {\rm gcd}(j,m)=1}}
\lr{
x-e^{\frac{2\pi j}{m}i}
}
.
\end{align*}
It is well known that this polynomial is irreducible over the field of the rational numbers.
Moreover, for any prime number $p$ and $n\in\NM$, the following relations hold.
\begin{align*}
\Phi_p(x)=\sum_{j=0}^{p-1}x^j,
\qquad
\Phi_{p^n}(x)=\Phi_p(x^{p^{n}-1}).
\end{align*}
We define an integer coefficient polynomial $f(x)=\sum_{j=0}^{k}q_jx^j$ satisfying the assumption of Proposition \ref{sumq}, i.e.,
\begin{align*}
f(\omega)=\sum_{j=0}^{k-1}q_j\omega^j=0,
\end{align*}
where $\omega=e^{\frac{2\pi}{k}i}$.
Here, if $f(x)$ is not divided by $\Phi_k(x)$, then the Euclidean method gives that there exist the polynomials with rational number coefficients $A(x)$ and $B(x)$ satisfying
\begin{align*}
f(x)A(x)+\Phi_k(x)B(x)=1.
\end{align*}
However, in the case of $x=\omega$, above equation does not hold.
Therefore, $f(x)$ is divided by $\Phi_k(x)$ and expressed as 
\begin{align}
\label{prove1}
f(x)=C(x)\Phi_k(x),
\end{align}
with an integer polynomial $C(x)$.
In particular, $f(x)=q_{p-1}\Phi_p(x)$ for $k=p$.
Noting that $f(1)=\sum_{j=0}^{k-1}q_j$ and $\Phi_k(1)=\Phi_p(1)=p$, Eq. (\ref{prove1}) gives
\begin{align*}
\sum_{j=0}^{k-1}q_j=p\,C(1).
\end{align*}
The desired conclusion is given by above relation.

\section{Examples}
In this section, we present some examples.
To count the number of paths, we introduce the adjacency matrix on a graph $G$ which does not have multiedges:
\begin{align*}
A_G=\lr{a_{x,y}}_{x,y=0,1,\ldots,N-1},
\qquad
a_{x,y}=
\begin{cases}
0\qquad (x,y)\not\in D(G)
\\
1\qquad (x,y)\in D(G)
\end{cases}.
\end{align*}
In general, when the graph has multiedges, the component of $A_G$ is defined by $m((x,y))$ instead of $1$ in the above definition.
Here, $m(e)$ denotes the multiplicity of the arc $e$.
By definition, it would be natural to give the relation $\left( A_G^t \right)_{x,y}=W_t(x,y)$.

\subsection{Complete graph $K_N$}
The complete graph $K_N$ with $N=p^n+1$ defined by $V(K_N)=[N]$ and $D(K_N)=\{(x,y)\ |\ x\neq y\}$ is an $N-1(=p^n)$ regular graph.
From now on, we show the following proposition.
\vspace*{12pt}
\begin{pro}
For $N=p^n+1$ with $p\in\PM$ and $n\in\NM$, the FQW on $K_N$ is not periodic.
\end{pro}
\vspace*{12pt}
\noindent
First, the components of its adjacency matrix $A_{K_N}$ are given by
\begin{align*}
a_{x,y}=
\begin{cases}
1\qquad (x\neq y)
\\
0\qquad (x= y)
\end{cases}.
\end{align*}
We can easily confirm that the components of $A_{K_N}^t$ can be expressed as $a^{(t)}$ and $b^{(t)}$ as follows.
\begin{align}
\label{At}
\lr{A^t_{K_N}}_{x,y}=
\begin{cases}
a^{(t)}\quad (x=y)
\\
b^{(t)}\quad (x\neq y)
\end{cases}
.
\end{align}
We now prove a statement, $a^{(t)}-b^{(t)}\equiv 1\ {\rm mod}\ p$ for any $t\in\NM$. 
Then, our main result shows that the FQW on $K_N$ is not periodic, since Theorem \ref{mainthm} guarantees  $a^{(T)}-b^{(T)}\equiv\ 0\ {\rm mod}\ p$ if this FQW has a finite period $T$.
By Eq. (\ref{At}), we have
\begin{align*}
a^{(t+1)}=(N-1)b^{(t)},\qquad b^{(t+1)}=a^{(t)}+(N-2)b^{(t)}.
\end{align*}
From these relations, $a^{(t+1)}-b^{(t+1)}=-a^{(t)}+b^{(t)}$ is given.
Hence, if $a^{(t)}-b^{(t)}\equiv 1\ {\rm mod}\ p$ holds, then $a^{(t+1)}-b^{(t+1)}\equiv 1\ {\rm mod}\ p$ also holds.
Noting that $a^{(1)}-b^{(1)}=1$, by induction on $t\geq 1$, we concludes the desired statement $a^{(t)}-b^{(t)}\equiv 1\ {\rm mod}\ p$ for any $t\in\NM$.

\subsection{Cycle graph $C_N$ with self loops}
The cycle graph with self loops which has $N(>2)$ vertices, $C_N^{(l)}$, is a 3-regular graph. 
We put $V\lr{C_N^{(l)}}=[N]$ and $D\lr{C_N^{(l)}}=\{(x,y)\ |\ x=y,\ x\equiv y\pm 1\ {\rm mod}\ N \}$.
Here, the components of its adjacency matrix $A$ are
\begin{align*}
a_{x,y}=
\begin{cases}
1\qquad x=y,\ x\equiv y\pm 1\ {\rm mod}\ N
\\
0\qquad otherwise
\end{cases}.
\end{align*}
The matrix form of $A_{C_N^{(l)}}$ is expressed as
\begin{align*}
A_{C_N^{(l)}}=
\begin{bmatrix}
1 & 1 & 0 & 0 & 0 & \cdots & 0 & 0 & 1\\
1 & 1 & 1 & 0 & 0 & \cdots & 0 & 0 & 0\\
0 & 1 & 1 & 1 & 0 & \cdots & 0 & 0 & 0\\
0 & 0 & 1 & 1 & 1 & \cdots & 0 & 0 & 0\\[-5pt]
0 & 0 & 0 & 1 & 1 & \ddots & 0 & 0 & 0\\[-5pt]
\vdots & \vdots & \vdots & \vdots & \ddots & \ddots & \ddots & \vdots & \vdots\\[-5pt]
0 & 0 & 0 & 0 & 0 & \ddots & 1 & 1 & 0\\
0 & 0 & 0 & 0 & 0 & \cdots & 1 & 1 & 1\\
1 & 0 & 0 & 0 & 0 & \cdots & 0 & 1 & 1\\
\end{bmatrix}.
\end{align*}
By using Theorem \ref{mainthm}, we can show the following proposition.
\vspace*{12pt}
\begin{pro}
For $N\neq 3^n\ (N>2)$ with $n\in\NM$, the FQW on $C_N^{(l)}$ is not periodic.
\end{pro}
\vspace*{12pt}
\noindent
In order to prove this proposition, we introduce
\begin{align*}
a_{i}^{(t)}=\lr{A_{C_N^{(l)}}^t}_{i,0}\ {\rm mod}\ 3.
\end{align*}
Here, the following lemma holds.
\vspace*{12pt}
\begin{lem}
\label{expro}
If there exist $m,t\in\ZMP$ such that $a^{(t)}_{i+3^m}=a^{(t)}_{i}$ for any $i\in [N]$, then $a^{(t-1)}_{i+3^{m+1}}=a^{(t-1)}_{i}$ for any $i\in[N]$.
\end{lem}
\vspace*{12pt}
This lemma implies that the FQW on $C_N^{(l)}\ (N\neq 3^n)$ is not periodic.
Hence, if the FQW has a finite period $T$, then $a^{(T)}_{i+3^m}=0$ for any $i\in[N]$.
From Lemma \ref{expro}, we see that $a^{(1)}_{i+3^{m+T-1}}=a^{(1)}_{i}$ for any $i\in[N]$.
However, it can not be satisfied, since $N\neq 3^n$.\\

\noindent
{\it Proof.}\quad
We assume that there exists $m,t\in\NM$ such that $a^{(t)}_{i+3^m}$ is independent of $i$.
First, we calculate the following equation.
\begin{align*}
\nonumber
\sum_{j=0}^{3^m-1}\left(
a_{i+1+3j}^{(t)}-a_{i+3j}^{(t)}
\right)
&=\sum_{j=0}^{3^{m-1}-1}
{\bigg \{}
\lr{
a_{i+1+3j}^{(t)}+a_{i+1+3j+3^m}^{(t)}+a_{i+1+3j+2\cdot 3^m}^{(t)}
}
\\
\nonumber
&
\hspace{2.5cm}
-
\lr{
a_{i+3j}^{(t)}+a_{i+3j+3^m}^{(t)}+a_{i+3j+2\cdot 3^m}^{(t)
}
}
{\bigg \}}
\\
&
=
3
\sum_{j=0}^{3^{m-1}-1}
\left(
a_{i+1+3j}^{(t)}-a_{i+3j}^{(t)}
\right)
.
\end{align*}
Here, the second equality is obtained by the above mentioned assumption.
Thus, we have
\begin{align}
\label{ex1}
\sum_{j=0}^{3^m-1}\left(
a_{i+1+3j}^{(t)}-a_{i+3j}^{(t)}
\right)
\equiv 0\ {\rm mod}\ 3
.
\end{align}
Moreover, by definition, we see that
\begin{align*}
a_i^{(t)}=a_{i-1}^{(t-1)}+a_{i}^{(t-1)}+a_{i+1}^{(t-1)}
\ {\rm mod}\ 3.
\end{align*}
From this relation, the left hand side of Eq. (\ref{ex1}) is expressed as
\begin{align}
\label{ex2}
\sum_{j=0}^{3^m-1}\left(
a_{i+1+3j}^{(t)}-a_{i+3j}^{(t)}
\right)
=
\sum_{j=0}^{3^m-1}\left(
a_{i+2+3j}^{(t-1)}-a_{i-1+3j}^{(t-1)}
\right)
\ {\rm mod}\ 3
\ 
=
a_{i-1+3^{m+1}}^{(t-1)}-a_{i-1}^{(t-1)}\ {\rm mod}\ 3.
\end{align}
Combining Eq. (\ref{ex1}) with Eq. (\ref{ex2}), we get the following desired conclusion.
\begin{align*}
a_{i-1+3^{m+1}}^{(t-1)}=a_{i-1}^{(t-1)}.
\end{align*}

\subsection{Hypercube $\QC_d$}
The hypercube with dimension $d$, $\QC_d$, is a $d$-regular graph and the number of its vertices is $2^d$.
The series of hypercubes is constructed by the following recursion. 
\begin{align*}
\QC_{d+1}=\QC_{d}\Box K_2\quad (\QC_{1}=K_2).
\end{align*}
Here, $X\Box Y$ means the cartesian product of graphs $X$ and $Y$.
\vspace*{12pt}
\begin{pro}
For $d=p^n\ (p\neq 2)$ with $p\in\PM$ and $n\in\NM$, the FQW on $\QC_d$ is not periodic.
\end{pro}
\vspace*{12pt}
First, the adjacency matrix of $\QC_{d}$ is also given by the recursion as follows.
\begin{align}
\label{recursion}
A_{\QC_{d+1}}=A_{\QC_{d}}\otimes I_{2} + I_{2^d}\otimes A_{K_2}.
\end{align}
Moreover, the adjacency matrix of $K_2(=\QC_1)$ and its eigensystems are given as
\begin{align*}
A_{K_2}(=A_{\QC_1})=
\begin{bmatrix}
0 & 1 \\ 1 & 0
\end{bmatrix}
,\quad
A_{K_2}\ket{u_{\pm 1}}=(\pm 1)\ket{u_{\pm 1}},
\end{align*}
where $\ket{u_{\pm 1}}=\,{}^T[\, 1\ \pm 1\,]$. 
By using these notations, Eq.(\ref{recursion}) gives eigensystems of $A_{\QC_{d}}$.
For $L=\{-1,\ +1\}^d$,
each eigenvalue is induced by any $l=(l_1,\,l_2,\,\ldots ,\,l_d)\in L$ as
\begin{align*}
\lambda_l=\sum_{i=1}^d l_i.
\end{align*}
Furthermore, the eigenvector associated with $\lambda_l$ is obtained by
\begin{align*}
\ket{v_l}=\bigotimes_{i=1}^d \ket{u_{l_j}}.
\end{align*}
Here, $||\ket{v_l}||^2=d$ and, the eigenvectors $\ket{v_l}$ and $\ket{v_{l'}}$ are orthogonal for $l,l'\in L\ (l\neq l')$.
Hence, a spectral decomposition of $A_{\QC_{d}}$ is expressed as
\begin{align*}
A_{\QC_{d}}=\sum_{l\in L}\lambda_l\times\frac{1}{d}\ket{v_l}\bra{v_l}.
\end{align*}
Put $k=(k_1,\,k_2,\,\ldots ,\,k_d)\in L$ with
\begin{align*}
k_i=
\begin{cases}
+1\quad (1\leq i\leq m+1)
\\
-1\quad (N+1< i \leq 2m+1)
\end{cases},
\end{align*}
where $m\in\NM$ with $2m+1=d$.
Then, we see that eigenvalue $\lambda_k$ equals to $1$ and for any $t\in\NM$, 
\begin{align}
\label{vk}
A_{\QC_{d}}^t\ket{v_k}=\sum_{l\in L}(\lambda_l)^t\times\frac{1}{d}\ket{v_l}\braket{v_l}{v_k}=\ket{v_k}.
\end{align}
Here, if the FQW on $\QC_d$ has a finite period $T$, then any component of $A_{\QC_d}^T \ket{v_l}$ is divisible by $d$, since $\ket{v_l}$ is an integer-valued vector and Theorem \ref{mainthm} guarantees that any component of $A_{\QC_d}^T$ can be divided by $d$.
However, it contradicts to Eq.(\ref{vk}) with $t=T$, since each component of $\ket{v_k}$ is $\pm 1$.

\section{Summary}
In this paper, we gave a necessary condition for the FQW on regular graphs to have the finite period.
As an application of our result (Theorem \ref{mainthm}), we proved that the FQW do not have any finite period for some classes of regular graphs like complete graph $K_N$ with $N=p^n+1$, cycle graph with self loops $C_N^{(l)}$ with $N\neq 3^n\ (N>2)$, and hypercube $\QC_d$ with $d=p^n\ (p\neq 2)$, where $p\in\PM$ and $n\in\NM$.

One of the interesting future problems would be to apply our result to other graphs, e.g., strongly regular graphs, the Hamming graphs, and the Moore graphs.

\noindent
\\
{\bf Acknowledgement.} The author would like to thank Norio Konno for useful comments.

\par
\
\par

\begin{small}
\bibliographystyle{jplain}

\end{small}

\end{document}